\documentclass[twoside,a4paper,11pt]{sea9}
\usepackage{graphicx}
\usepackage{hyperref}
\usepackage{color}
\begin{document}
\pagenumbering{arabic}
\pagestyle{myheadings}
\thispagestyle{empty}
{\flushleft\includegraphics[width=\textwidth,bb=58 650 590 680]{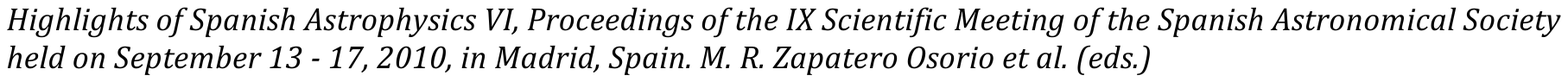}}
\vspace*{0.2cm}
\begin{flushleft}
{\bf {\LARGE
%
The Observatorio del Teide welcomes SONG: The Stellar Observations Network Group%
}\\
\vspace*{1cm}
%
O.~L.~Creevey$^{1,2}$,
F.~Grundahl$^{3}$
P.~L.~Pall\'e$^{1,2}$, 
U.~Gr{\aa}e  J{\o}rgensen$^{4,5}$, \\ 
J.~A.~Belmonte$^{1,2}$,
J.~Christensen-Dalsgaard$^3$, 
S.~Frandsen$^3$, 
H.~Kjeldsen$^3$ and 
P.~Kj{\ae}rgaard Rasmussen$^4$ 
%
}\\
\vspace*{0.5cm}
%
$^{1}$
Instituto de Astrof\'{i}sica de Canarias, C/ V\'ia L\'actea s/n,
E-38200 Tenerife, Spain.\\
$^{2}$Universidad de La Laguna, Avda. Astrof\'isico 
Francisco S\'anchez s/n, 38206 La Laguna, Tenerife, Spain.\\
$^{3}$
Department of Physics and Astronomy, Aarhus University, 
         Ny Munkegade, 8000 Aarhus C, Denmark\\
$^{4}$
Niels Bohr Institute, University of Copenhagen, Juliane 
         Maries Vej 30, 
         2100 Copenhagen, Denmark\\
$^{5}$
Centre for Star and Planet Formation, University of Copenhagen, 
         Geological Museum, {\O}ster Voldgade 5-7, 1350 Copenhagen, Denmark\\
         
%
\end{flushleft}
%
\markboth{
The Observatorio del Teide welcomes SONG
}{ 
%
Creevey et al.
%
}
\thispagestyle{empty}
\vspace*{0.4cm}
\begin{minipage}[l]{0.09\textwidth}
\ 
\end{minipage}
\begin{minipage}[r]{0.9\textwidth}
\vspace{1cm}
\section*{Abstract}{\small
%

The Stellar Observations Network Group (SONG) is an international
network project aiming to place eight 1-m robotic telescopes around the globe,
with the primary objectives of studying stellar oscillations and planets
using ultra-precision radial velocity measurements.
The prototype of SONG will be installed and running at the Observatorio
del Teide by Summer 2011. 
In these proceedings we present the project, primary scientific objectives, and instrument, and discuss
the observing possibilities for the Spanish community.

%
\normalsize}
\end{minipage}
%
%
%
\section{Introduction \label{intro}}

The Stellar Observations Network Group (SONG) is an 
initiative to design and build a global network of small 
telescopes. The goal of SONG is to become a key facility in 
both asteroseismology and planet search research 
programmes, a facility that provides state-of-the-art data of a 
quality that could not be achieved by use of any other space-based 
or ground-based facility \cite{jcd06,gru06,gru08a,gru08b,gru09}.  

SONG is proposed to have a total of eight nodes,
four located in the northern hemisphere and four in the south.
By placing the telescopes at roughly equally-spaced 
longitudes, long-term nearly continuous observations
can be obtained.
Each telescope 
has an aperture of 1m and will be equipped with
   a high resolution spectrograph for measuring very precise doppler
   velocities and dual--color lucky-imaging cameras for photometry
   of faint stars in crowded fields.

The prototype of SONG has been fully financed and 
a site on the Observatorio del Teide is being
prepared for its installation.

\section{Scientific Goals}
The primary scientific goals of SONG are to conduct
asteroseismology for bright solar-type stars and to 
characterise extra-solar planetary systems.
SONG will be open to collaboration and proposals 
for any other scientific
topic are welcome, including long-term continuous observations.

\subsection{Asteroseismology}
Asteroseismology is the study of the interiors of stars using 
observations of stellar pulsations  (e.g. \cite{bg94,jcd05,aer10}).
The stellar oscillations can be determined with extremely high
precision ($<0.1$\%) so if they can be measured, they 
provide strict constraints for stellar modelling  
\cite{ulr70,ls71,jcd07,ste09,met10,deh10,dm10,vg10,cb11}.
The frequencies are determined by the sound-speed
profile across the star, and so they are sensitive 
to the mass and size of the star, internal structure and rotation.
The amplitudes and lifetimes of the modes are 
sensitive to near-surface physics, including 
convective dynamics.

The goals of asteroseismology depend on the observations that
are available for a star.
One of the primary aims is the characterise global
stellar properties, for example, determine the mass, radius, 
mean density, age, 
luminosity, and chemical composition.
If we have a stellar laboratory that is well-characterised,
then we can go a step further and 
investigate detailed internal structure and 
dynamics of stars.
Asteroseismology aims to improve our understanding 
of the physics of stellar interiors, and thus aims to 
improve stellar models.


The first power spectrum showing the Sun's five-minute oscillations appeared
in the literature roughly thirty years ago \cite{gre80}.  
Today, the power spectra for other stars far exceeds the 
quality of those first spectra.
We have been able to measure oscillations
 using ground-based spectrographs for over 20 such stars (e.g. \cite{bk03,bru10}).
In figure 10 of \cite{bk03}, they show how the 
acoustic power varies with stellar properties.
 Detected power is 
found at increasing frequencies as the acoustic
 cavities (the sizes of the stars) shrink,
while the amplitude of oscillations increases as the acoustic cavity grows \cite{der09}.
 The size of the acoustic cavity depends mainly on the radius (mass)
 and the age (chemical profile) of the star.
 
So, why do we need a ground-based network measuring radial velocities?
To date, we have vast quantities of data for solar-like stars on fainter stars from the 
CoRoT and Kepler missions (e.g. \cite{app08,mos09,bed10,hek10}).  
Because the stars are fainter, this implies that the oscillation amplitudes are 
even more difficult to detect.  
The most recent observations of such stars have been restricted to 
oscillations in red giants, where the amplitudes are much more
detectable.  These satellite missions measure in photometry and although
we can observe many stars simultaneously, photometry is sensitive only 
to the modes of degree 0, 1, and 2.

Radial velocity measurements can reach precisions typically below 1m/s.
The tiny amplitudes of oscillations in 
main sequence stars can thus be observed, 
and mode degrees $\ell < 4 $ will be detected.
SONG will observe only bright targets ($V < 6$) and these  
can be well-characterised with complementary observations, such 
as interferometry and spectroscopy.   The additional observations
constrain the model parameters, so that the frequencies can be used uniquely 
to learn about the stellar interior.

\subsection{Extra-solar planets}
The characterisation of extra-solar planetary systems will be done using 
radial velocity observations and gravitational micro-lensing. 
The radial velocity observations will allow us to identify and characterise masses of planets 
much smaller than those from traditional radial velocity searches. 
Simulations of SONG radial velocity data based on convolving available SOHO solar data, 
indicates that we will be able to identify Earth-mass planets orbiting solar type stars 
with periods shorter than 8 days, Mars-mass planets in orbits shorter than 1 day, 
and Earth-mass planets in orbits shorter than 20 days. 
The major SONG exoplanetary effort will, however, be through microlensing observations, 
and the design of the lucky imaging camera is optimized to make such observations most efficient.  
The microlensing time series at SONG will be sensitive to analogues of all of the planets in our 
own solar system (except Mercury), 
including planets as small as Mars in terrestrial-like orbits around solar type stars.

When two stars pass very close to one another on the sky, the gravitational field from the foreground star (the lensing star) will magnify the light from the background star (the source star). SONG will follow the magnified light curve of many hundreds of such stellar lensing events per year. If the lensing star is orbited by a planet, the gravitational field from the planet can cause an asymmetry in the magnified light curve. Analyses of the form and magnitude of such asymmetries give us information about the planetary mass and orbit.  If standard theories of planet formation are right, we will expect the SONG network to discover of the order of 100 new terrestrial-like extrasolar planets per year. Also the seemingly rare Saturn-Uranus-Neptune analogues in Saturn-Uranus-Neptune like orbits are within the reach of the SONG microlensing programme.

\section{Design \& Characteristics}
\subsection{Telescope}
\begin{figure}
  \center{\includegraphics[height = 7cm]{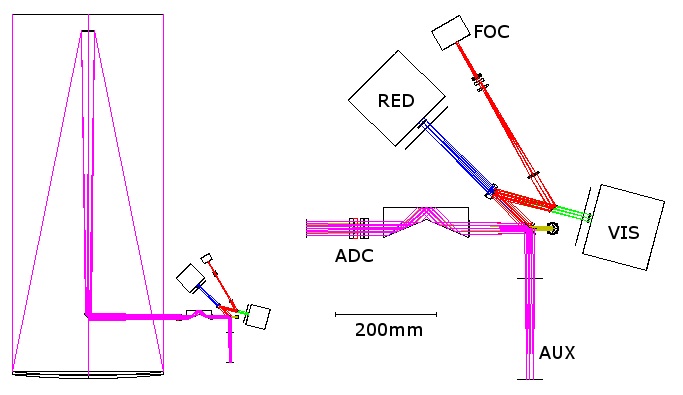}}
  \caption{Light path.
  Once the light reaches one of Nasmyth focii, 
  the light goes through the ADCs, the de-rotator and then to the three-position wheel:
  either to the Coude path (yellow path), 
  to the two lucky-imaging cameras (red path), or to an auxiliary position.\label{fig:one}}
  \end{figure}

The telescope for SONG has an aperture of 1m and is equipped with a
  coude path which feeds the light to the high resolution spectrograph, 
  and two nasmyth foci used for the lucky imagers (see Fig.~\ref{fig:one}). 
  It is housed in a
  modified Ash dome of $\approx$5m diameter. To ensure a good image
  quality the thin (5cm) primary mirror is equipped with active optics. 
  This will ensure optimal performance for the lucky-imaging observations
  of microlensing fields. Two nasmyth foci are available (initally only
  one will have instruments) via a computer controlled tertiary mirror. 
  The primary nasmyth focus is equipped with an optical image derotator
  of the Abbe-K\"{o}nig type and an atmospheric dispersion corrector (ADC).
  Since the telescope must be able to acquire objects automatically the
  blind pointing will be better than 5 arcseconds. Slew speeds can be up
  to 20 degrees per second.

\subsection{Nasmyth Instrumentation}
\begin{figure}
  \center{\includegraphics[height = 7cm,width=8cm]{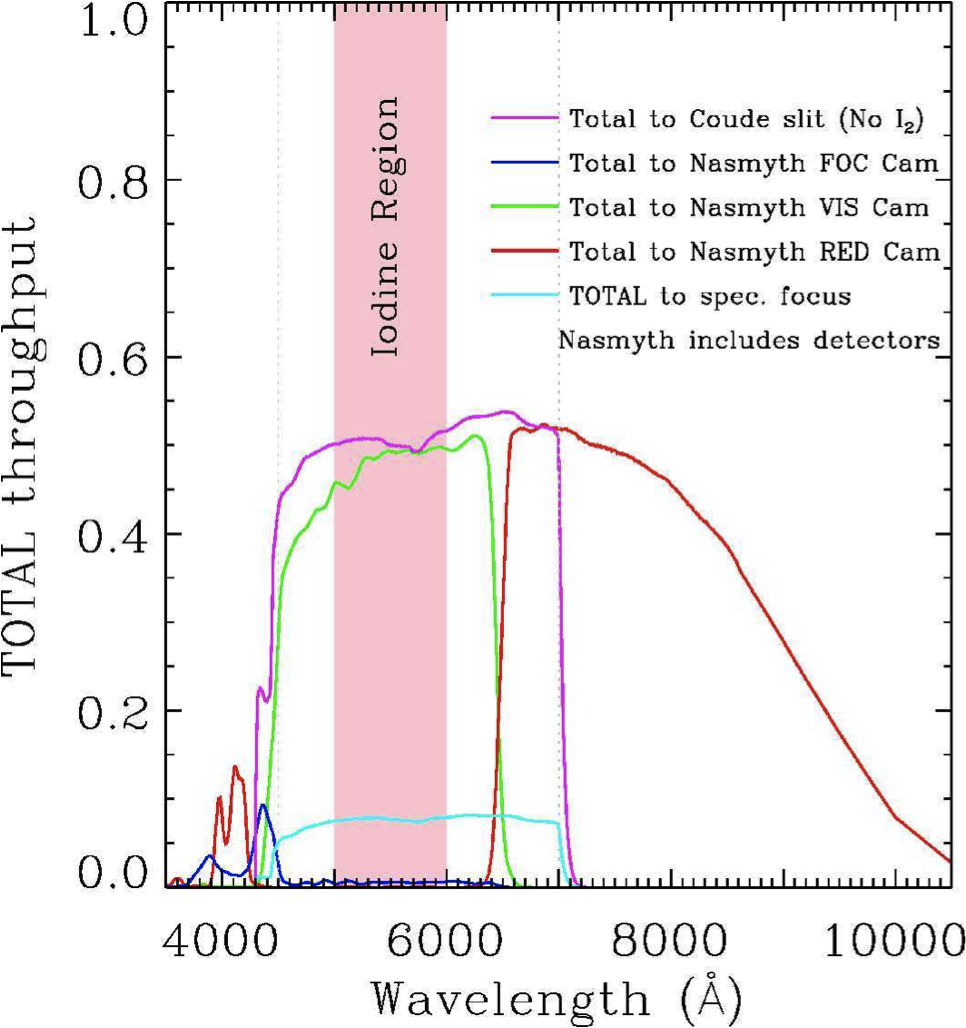}}
  \caption{The observable spectral range with SONG.\label{fig:spectrum}}
  \end{figure}

 For microlensing observations one of the nasmyth foci will be equipped
  with two lucky-imaging cameras to allow dual-color, 
  simultaneous, imaging in a visible and red channel, with a wavelength split
  at 650nm (see Fig.~\ref{fig:spectrum}). Since the field of view required for the photometric observations
  is small we have designed the instrument with a sampling of 0.09 arcsecond
  per pixel, which allows to sample the diffraction limit at 800nm with 
  two pixels. 
  We have adopted the lucky-imaging method in order to obtain high-resolution
  images for the photometry. The microlensing fields towards the Galactic 
  Bulge are typically very crowded and for such fields a high spatial 
  resolution is clearly advantageous. 
  Lucky imaging is already well developed at OT (see  
   http://spie.org/x648.html?product\_id=788834) and has demonstrated that 
  near diffraction limited performance for a 1.5m telescope at OT is possible. 
  Due to the smaller aperture of the SONG telescope we can expect that a
  large fraction of the images can be used to ensure images of 0.5 arcsecond
  for a substantial part of the available time. 
  
\subsection{Spectrograph}
\begin{figure}
  \center{\includegraphics[height = 6cm]{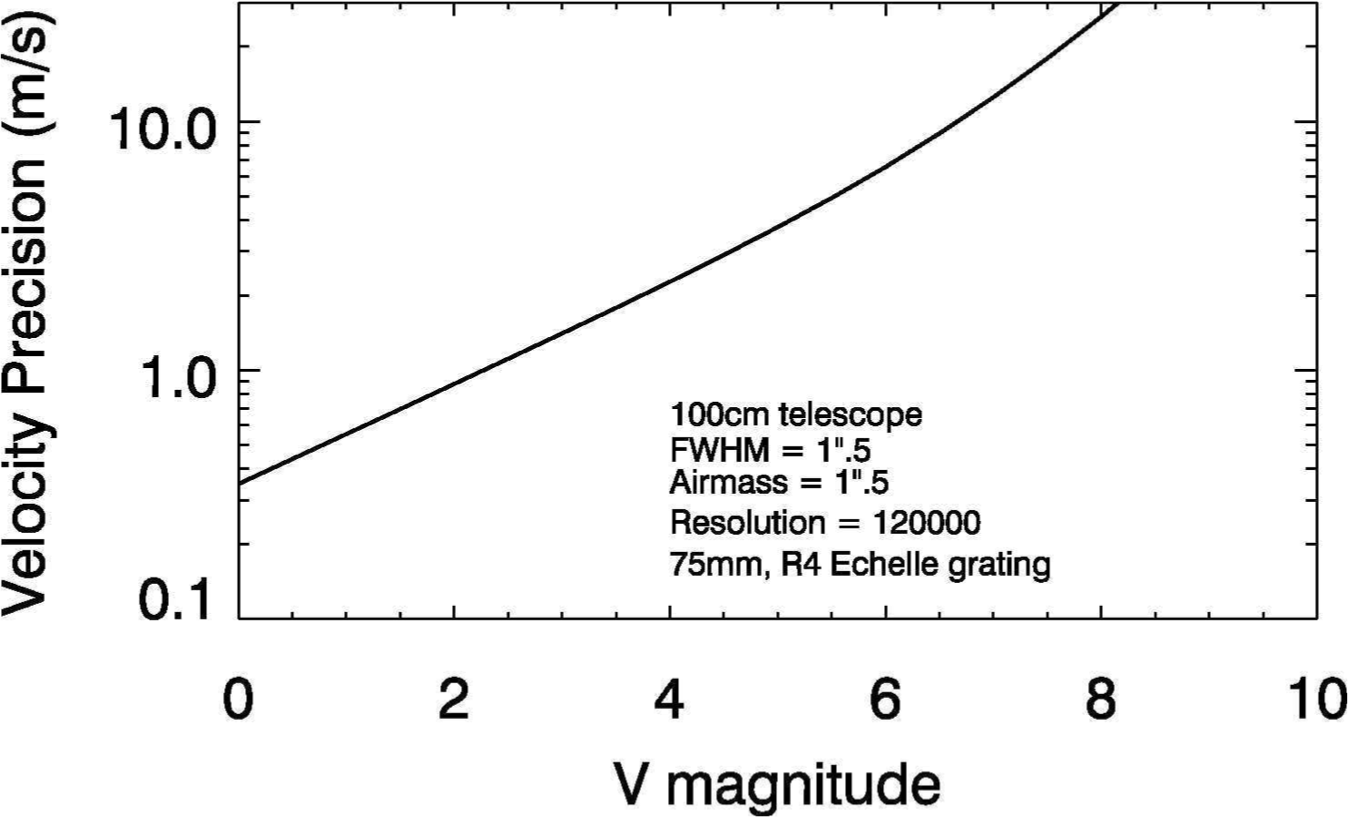}}
  \caption{Radial velocity precision as a function of stellar $V$ magnitude, for a spectral resolution
 of 120,000. \label{fig:velpre}}
  \end{figure}

The asteroseismic observations require very precise doppler velocities
  of the stars under study. To enable such measurements the spectrograph
  is designed (by Paolo Spano) for high resolution and efficiency. The 
  aim is to be able to measure velocities with a precision of 1m/s for the
  brightest ($V<2$) stars (see Fig. \ref{fig:velpre}). 

  Spectral resolutions between 60,000 and 180,000 are available, 
  although the two pixel sampling limit is at 120,000. The spectrograph
  in located in a temperature controlled box at a coude focus in the 
  container next to the telescope pier. For wavelength reference we employ
  an iodine cell. All optics in the system (coude path + spectrograph) have
  been optimized for the 480--680nm region to provide a high throughput. 

  The spectrograph beam diameter is 75mm, and an R4 echelle is used, and the
  resolution of 120,000 is achieved with a 1 arcsecond slit. In order to 
  provide a stable illumination of the slit the light is fed to the 
  spectrograph via a fast tip/tilt mirror, allowing corrections up to 100Hz. 
  Furthermore, we have a camera for monitoring the location of the telescope
  pupil (to make sure the illumination of the grating is not changing), and
  one of the coude mirrors can be used to control the pupil location if it
  changes in time. 
  
  The iodine cell, flat field lamp, ThAr lamp and guiding system is 
  mounted in a separate pre-slit unit before the light enters the 
  temperature stabilized spectrograph box. Note, that the iodine cell can
  be removed from the light path, allowing the spectrograph to be used
  in a ``normal'' fashion, as requested by the user. 
\section{Operations \& Current Status}
There are currently (December 2010) many activities ongoing at Aarhus and 
Copenhagen Universities, and at the Observatorio del Teide.
 The mechanical parts for the instruments are nearly
completed, with only minor components missing, and the assembly and integration
of the spectrograph is starting. A test container is installed in Aarhus, 
where all items will be mounted and system integration and testing will be performed, before
the instruments are shipped to Tenerife. 
The software for controlling the prototype, and ultimately the network is
being developed, and the setup for automatic execution of observations and
copying of data to archives and databases is ready. 
The site at Observatorio del Teide is being prepared, and installation of the
dome support structure and container is scheduled for early 2011, followed by
an extensive testing period lasting until the end of 2011. 

Arrangements have already been made to construct a second SONG node in China, and this should be
operating by Summer 2012.
The much improved spectral window will allow us to obtain data with much better precision on the 
individual frequencies.

In Spain, the SONG activities are being led by 
the research group at the Instituto de Astrof\'isica de Canarias.
Interest in using the SONG telescope should be directed towards any of these authors.
The scheduling of the observations has not been finalised, and input is requested from
the full scientific community.  
Ideally each telescope node (or nodes) will dedicate a block of time to observing one target for
asteroseismology, then a few months will be exclusively used for planet-searching.
A third block of observations can be granted to exciting proposals that require these telescopes 
characteristics, and in this sense, the telescope is open to all areas of scientific interest.
It is also possible to propose single observations, or periodic observations, such as, a single
spectrum at the beginning of each night during a certain period of time.
All proposals are welcome and will be considered.
More information can be found at \url{http://astro.phys.au.dk/SONG/}.

%
%
\small  
%
\section*{Acknowledgments}   
%
The authors gratefully acknowledge generous financial support for the SONG
project from the Villum Fonden, Carlsbergfondet and The Danish Council for 
Independent Research $|$ Natural Sciences. 
Funding for the local construction costs were covered by the
Department of Research of the Instituto de Astrof\'isica de Canarias.

%

%
\end{document}